\documentclass[a4paper]{article}
\usepackage[affil-it]{authblk}
\usepackage{graphicx}
\usepackage[authoryear]{natbib}
\usepackage[top=2cm,bottom=2cm,left=2.5cm,right=2.5cm]{geometry}
\usepackage{setspace}
\usepackage{color}
\usepackage{caption}
\usepackage{subcaption}
\usepackage{tikz}

\title{Detailed comparative study and a mechanistic model of resuspension of spherical particles from rough and smooth surfaces}
\author{Ron Shnapp, Alex Liberzon%
\thanks{$^*$ Electronic address: \texttt{alexlib@eng.tau.ac.il}; Corresponding author}}
\affil{Turbulence Structure Laboratory, School of Mechanical Engineering, \\
Tel Aviv University, Tel Aviv 69978, Israel}

\date{Dated: \today}

\begin{document}
\maketitle

\begin{abstract}

Resuspension of solid particles by a tornado-like vortex from surfaces of different roughness is studied using a three-dimensional particle tracking velocimetry (3D-PTV) method. By utilizing the three-dimensional information on
particle positions, velocities and accelerations before, during and after the resuspension (lift-off) event, we demonstrate that the resuspension efficiency is significantly higher from the rough surface, and propose a  mechanistic model of this peculiar effect. The results indicate that for all Reynolds numbers tested, the resuspension rate, as well as particle velocities and accelerations, are higher over the rough surface, as compared to the smooth counterpart. The results and the model can help to improve modeling and analysis of resuspension rates in engineering and environmental applications.

\end{abstract}

\section{Introduction}

Particle resuspension is the process in which a submerged particle is being detached from a surface to the fluid medium above, after the break-up of the particle-surface bond. Resuspension is an ubiquitous process in many engineering
 and environmental applications, for instance in sediment transport \citep{Wu:2013}, powder handling processes \citep{Grzybowski:2007}, and studies of Martian dust devils \citep{Greeley:2004}. Previous studies have shown that resuspension is inherently related to the flow regimes near the wall, and in many cases may be related to turbulent flow velocity fluctuations or flow structures, such as sweeps and ejections \citep{henry_minier:2014}. The dependency of the resuspension phenomena on diverse flow regimes makes it hard to study in a general fashion. Therefore, the detailed study of resuspension demands a break-up of the mechanism as a whole into separate stages,
so in the future, the broad picture may be better understood.
 In this work we want to focus on the stage of the freely moving particle lift-off from smooth or rough surfaces.

Recent reviews of particle resuspension from surfaces by \citet{Ziskind:2006, henry_minier:2014} show that the complexity of the resuspension phenomena is caused by two inherent features: particle interaction with the surface to which it is attached, and particle interaction with the fluid to which it is exposed. It is also evident that these two features of the system are affected by each other, as the surface roughness affects the flow regime, and the fluid properties affect the bonds that the particles have with the surface (cohesive and adhesive forces). Furthermore, studies have shown different (often opposite) relations between the wall surface roughness, particle diameter and the efficiency of the resuspension process \citep[e.g.][among others]{nino:2003, YanbinJiang:2008, lee:2012, Barth:2014}.

Surface roughness effect on the resuspension rate can be modeled in many different ways and by different mechanisms. For instance, \cite{Henry:2012} in their model of re-entrainment coupled surface roughness with the effect of particle-surface adhesion. \cite{lee:2012} calculated a critical shear stress criterion for the initiation of particle movement, and determined that surface roughness may affect particle movement through the level of relative particle protrusion, or through a moment balance of hydrodynamical and resistive forces against an asperity. Using a channel airflow experiment, \cite{YanbinJiang:2008} demonstrated  that the effect surface roughness has on resuspension varies for particles of different sizes and for different scales of surface roughness. \cite{Hall:1988} measured and derived an expression for the lift force acting on a particle on smooth and rough surfaces, and found that the force can change by several orders of magnitude with surface roughness, and with the position of a particle relative to the roughness elements. In our study, we choose to focus on a single aspect of this diverse phenomena,
 namely on the way by which particle mobility (rolling or sliding motion along the surface) over the smooth and rough surfaces affects resuspension of relatively large spherical particles.

Many experimental methods have been introduced so far for measuring resuspension rates. The majority of studies were conducted through a wind tunnel or a duct flow with particles spread over the channel bed. The initial load of particles is measured and particles that leave the observation volume are counted, providing the fraction remaining \citep[][among others]{Ibrahim:2003, nino:2003, YanbinJiang:2008}. Other experimental methods intended to study resuspension under specific flows, such as the air flow generated by the foot during walking, or through porous medium, have also been introduced - examples can be found in a recent review by \cite{henry_minier:2014}. These methods allow to quantify the resuspension rates and test models of the resuspension problem at large. For our purpose of studying the basics of the resuspension mechanism, and focusing on a single major difference between the smooth and rough surfaces, an experiment with a confined flow and particle motion, along with the detailed three-dimensional measurements, is required.

In order to achieve a quasi-static state, a steady vortex flow type was chosen, mimicking several industrial and environmental applications such as mixers, bio-reactors, tornadoes or dust devils. The low pressure, found at the center of a vortex-like swirling flow, creates a ``suction'' effect that generates high lift forces over submerged bodies. As a result, vortex flows present higher resuspension rates at a low level of energy input to the system, as compared to the unidirectional boundary layer type of flows. Moreover, the low pressure zone at the vortex core keeps the initial group of particles within a observation volume, thus allowing high fidelity measurements and significant statistics based on long and detailed observations to be collected for relatively small groups of test particles.
Using a three-dimensional particle tracking velocimetry (3D-PTV) system, the particles' locations and velocities can be measured in time, and thus different aspects of their instantaneous and statistical behavior can be put under examination \citep[e.g.][]{Traugott:2011}. 


\section{Materials and methods}\label{sec:method}


\subsection{Experimental methods} 

\begin{figure}[ht]
   \centering
    \includegraphics[width=0.5\textwidth]{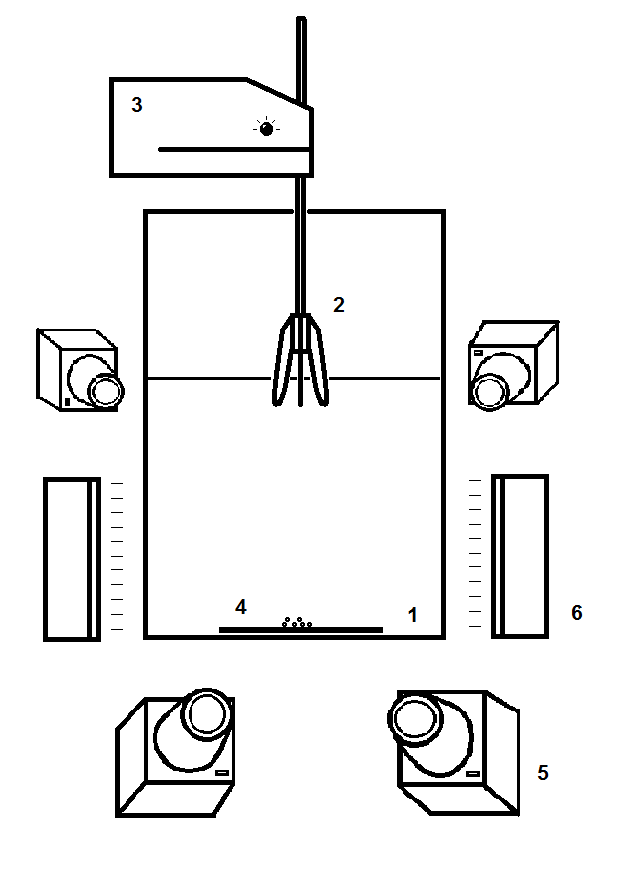}
    \caption{Schematic view of the experimental set up: (1) water tank  (2) four-blade rotor, (3) overhead stirrer, (4) smooth or rough bed with the particles, (5) four high speed digital cameras and (6) two LED line light sources \label{fig:setup}}
\end{figure}

The experimental set up is shown in figure~\ref{fig:setup}. A $300 \times 300 \times 400 mm^{3} $  glass tank filled with filter water at room temperature (density $\rho = 1000$ kg m$^{-3}$ and kinematic viscosity $\nu = 10^{-6}$ m$^2$s$^{-1}$) up to 230 mm height (1). A Four blade rotor (2) rotates on a shaft of a stirrer (3) equipped with an
angular velocity control (RD-03, MRC Inc.). At the tank bottom wall, the different roughness surfaces (4) were attached. Four high speed digital CMOS cameras (5) were placed around the tank, along with two LED lights (6). The digital video data was recorded to the RAM of a video recording unit and processed later using an open source particle tracking velocimetry software, OpenPTV (www.openptv.net). We tested four different angular velocities of 70, 100, 130 and 160 rpm and the bulk Reynolds number of this vortex type of flow is defined using the stirrer radius $R$, and the motor angular velocity $\omega$,  $Re = \omega R^2/\nu$. The corresponding Reynolds numbers tested are in the range of $13,000 \div 30,000$.  Each experiment was repeated at the steady state conditions, after the motor was running for at least five minutes with the particles in the tank, so the flow was allowed to reach a stable steady state vortex flow. After establishing the steady flow and resuspension conditions, a digital video sequence from different view angles was taken at a rate of 500 frames per second with the cameras focused at the bottom surface under the vortex core.

We used soda lime glass spherical particles $d_p = 800\, \mu$m in diameter and with the density of 2.5 g cm$^{-3}$. The ``smooth'' surface was made of a thin sheet of PVC with a layer of PVDF coating. This configuration yields a negligible surface roughness of $ Ra = 0.65\, \mu$m. The ``rough'' surface was made out of a sheet of 240 grit aluminium oxide sand paper, which yields a surface roughness of $Ra = 14.24\, \mu$m. Thus, relative to the particle diameter, the roughness $Ra/d_p$ is $0.81 \times 10^{-3}$ and $18.0 \times 10^{-3}$, respectively. Using 3D-PTV, we measured $v \approx 0.5$ m s$^{-1}$ at $y = 400\, \mu$m, a height of one particle radius above the floor, from which the friction velocity $u_{\tau}= 0.042$ m s$^{-1}$, and the thickness of the viscous sub-layer of $y = 120\ \mu$m can be estimated. Although this is a very rough estimate, it is clear that for both surface types, the surface roughness is significantly smaller than the viscous sub-layer thickness for the tested velocities.


Prior to the dynamic flow driven experiments, the effect of surface roughness on the incipient motion of the particles through rolling/sliding was tested using an inclined bed test. After a number of particles were positioned on the surface, an angle of inclination was gradually increased and the angle corresponding to the first motion was measured. The test results indicate that over the rough surface, the force required for the incipient motion (in this experiment we could not distinguish between a rolling or sliding motion) is about 3 times larger than for the smooth surface.


\subsection{Data reduction}\label{sec:data_reduction}

\begin{figure}[!ht]
    \includegraphics[width=0.8\textwidth]{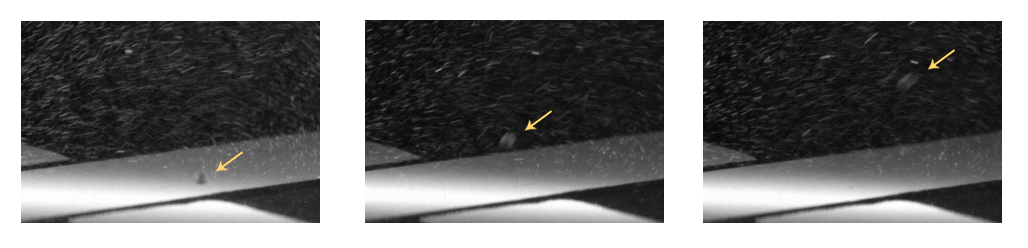}\\
    \includegraphics[width=1\textwidth]{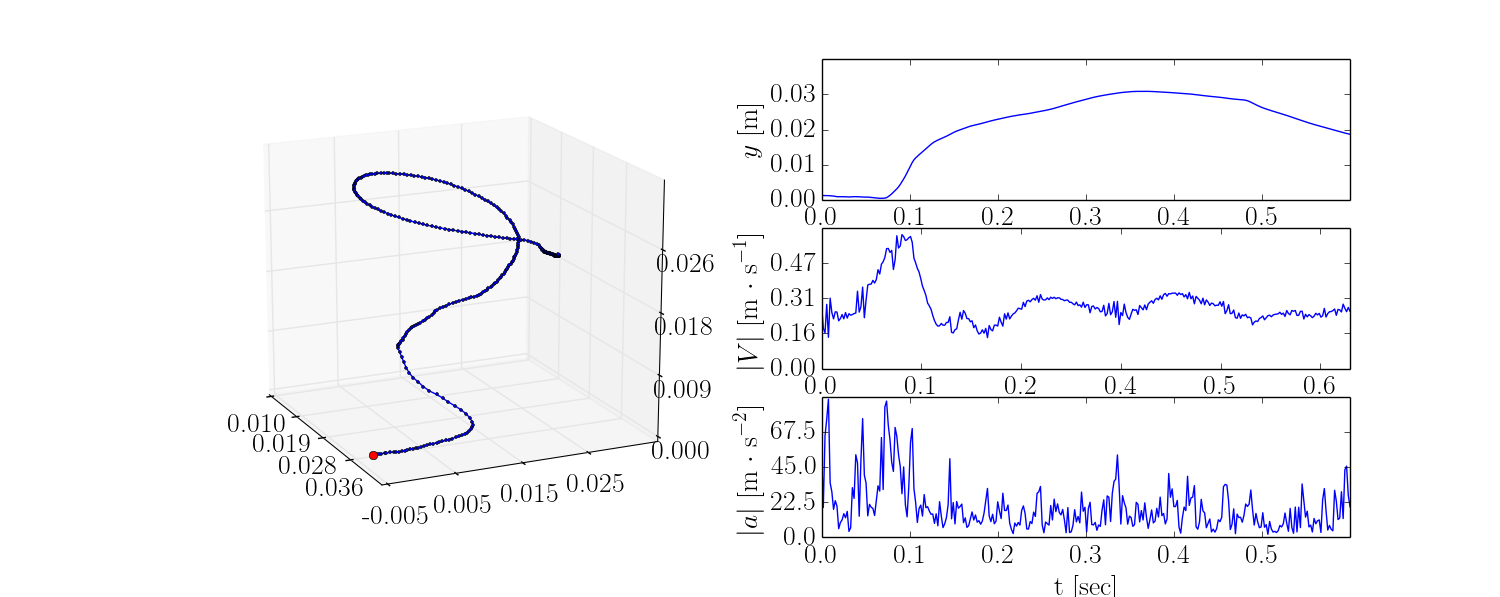}
    \caption{\label{fig:trajex} (top) Single particle at different stages of the resuspension event in a vortex flow; (bottom left) - isometric view of a single particle trajectory experiencing the resuspension event, (bottom right) - the height above the surface (top), the velocity (center) and acceleration (bottom) magnitude of the particle in time.}
\end{figure}

The captured videos were analyzed using OpenPTV, an open source particle tracking velocimetry software (www.openptv.net) that through a process of calibration and image processing, provides the 3D locations of the particle in each video frame \citep{Dracos:1996, Traugott:2011}. An example of the video frame, seen from one of the four cameras at different stages of a particle resuspension, is shown in the top panel of figure~\ref{fig:trajex}. The obtained dataset provides us with the information of particle position, velocity and acceleration in time as Lagrangian trajectories, shown for instance in the bottom panel of figure~\ref{fig:trajex}. There is a single particle trajectory during resuspension, shown in a three dimensional view, along with the quantitative information regarding the elevation above the wall $y$, the velocity magnitude $V = \sqrt{v_{x}^{2}+v_{y}^{2}+v_{z}^{2}}$,  and acceleration magnitude $a$, at different time instances. A sharp increase of the vertical velocity of the particle, which is still located on the
wall ($y \approx d_p/2$), is identified as a resuspension event, if at the following time interval the particle is lifted from the wall. Applying this condition, we can identify the particles freely moving along the bottom wall (either rolling or sliding) without vertical (against gravity direction) motion, and distinguish those from the particles that were resuspended.

%
%

\begin{figure}[ht]
	\centering
	\includegraphics[width=.8\linewidth]{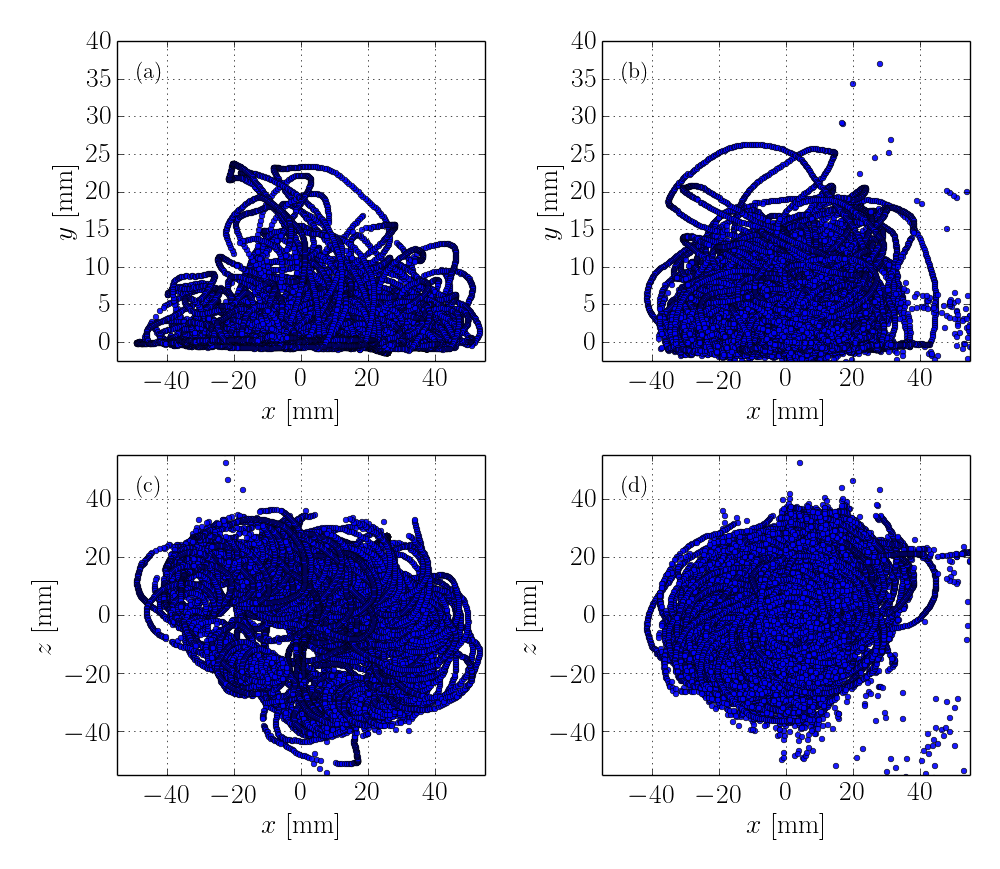}
	\caption{\label{fig:display} Particle position with 160 RPM angular velocity over (a) smooth surface side view and (b) top view (c) rough surface side view (d) top view}
\end{figure}

For the statistics, a larger dataset of the separate particles at different time instances, disregarding their Lagrangian trajectories, is used. An example of the graph demonstrating the gross picture of the particles at all times in two experimental cases is presented in figure~\ref{fig:display}. The two orthogonal views on the particle motion under a tornado-like vortex of the same angular velocity are shown. These views allow quantification of the probability density functions of particle horizontal (top panel) or height (bottom panel) positions, and the respective distributions of velocities or accelerations.

 An evolution of the particle motion in the given flow, shown in figure~\ref{fig:display} is the following. As the motor is started and the vortex starts to form, particles start moving on the bottom wall and gather at a location beneath the vortex core. After a certain time, when the vortex reaches a steady state, the particles behave in a repetitive cyclic manner of resuspension in a helical trajectory that is followed by a fall back to the surface (deposition). A more detailed observation reveals that the resuspension occurs mostly when the particle is close to the vortex core, while as the particles are swept up, their distance from the vortex core increases, their velocity decreases below their own settling velocity and they fall back to the bottom wall. This behavior of leaving the center of the vortex is due to the large centrifugal forces acting on the particles as they are being swept by the vortex in helical trajectories. This observation is evident in figures~\ref{fig:trajex} and \ref{fig:display}.

Integral quantities that characterize overall resuspension efficiency, such as the potential and kinetic energy of the particles (denoted $E_{p}$ and $E_{k}$, respectively) were calculated and presented here as normalized quantities per unit mass (in order to remove the effect of a slightly different number of particles at different runs).  The total energy was estimated accordingly, combining the potential and kinetic energy of the particle $i$ at position $y_i$ and velocity $V_i$ :

\begin{equation}\label{eq:E_T}
E_{T}=E_{P}+E_{K}= g\,y_{i}+\frac{V_{i}^{2}}{2}
\end{equation}
%
%
%

For the sake of resuspension efficiency comparison, the dimensionless resuspension rate $n^*$ was defined using a ratio of the number of resuspended particles $n_r$, to the total number of the particles in the frame, $n_T$:
\begin{equation}\label{eq:3}
n^* = 1-\frac{n_{r}}{n_{T}}
\end{equation}

In the next section, probability density functions of the resuspension rate, as well as the energy components will be presented. Moreover, we will show how the results help us to understand the differences between the resuspension rates from smooth and rough surfaces.


\section{Results}\label{sec:results}

\subsection{Energy distributions}

\begin{figure}[ht]
	\centering
	\includegraphics[width=1\linewidth]{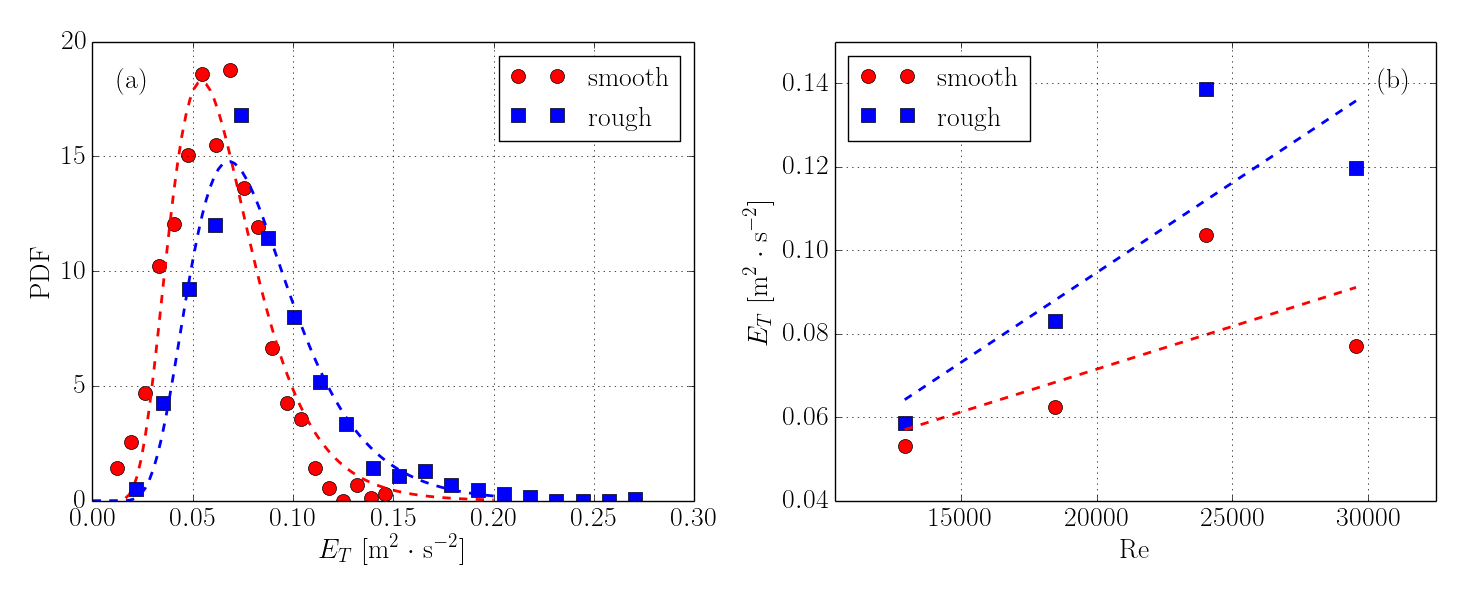}
	\caption{\label{fig:total_energy}(a) PDF for the particle kinetic and potential energy at 100 RPM motor velocities over the smooth and the rough surfaces. (b) mean particle energy values over the two surfaces with the changes in motor velocity }
\end{figure}

As a first measure for the resuspension efficiency of the system, the total energy ($E_T$) of the particles, calculated using Eq.~\ref{eq:E_T}, is considered. The probability density functions (obtained by the least-square fit to the measured histogram) are presented in figure \ref{fig:total_energy}(a), for the 100 rpm case, as an example. It can be seen that over the smooth surface, the distribution is denser and with a slightly lower energy level of the mode compared to the rough surface energy distribution. 
Moreover, although they seem to
behave in a similar manner, the particles above the rough surface exhibit a wider distribution of the total energy with a longer tail and with stronger probability of higher energy.
This behavior was seen for all cases studied here, and it is in agreement with the results of \cite{Soltani:1995} that showed a lower threshold of critical shear velocity for the resuspension of particles on a rough surface, with surface roughness much smaller that the particle diameter, as the case studied here.

The mean values of particle total energy distributions are plotted for the increasing vortex angular velocity in figure~\ref{fig:total_energy}(b). From the graph, a monotonic increase in total particle energy is visible in the range of 70 to 130 rpm or Reynolds of 13,000 to 30,000, and slightly reduced values for the highest velocity tested of $Re \approx 30,000$ (or 160 rpm). Although this decrease is slightly counter-intuitive (as one would expect to see a rise in particle energy for increasing values of the input energy), an explanation is due to the type of the flow and the description of particle motion given above: as the motor velocity is increased, the rise in angular velocity of the particles is related to stronger centrifugal forces acting on the particles, driving them farther away from the vortex core where the gravity forces are dominant and faster deposition.
 Thus, it can be derived that the vortex flow structure can not efficiently sustain particle suspension for all cases. This effect is evident in our results where the mean total particle energy reduced, while motor velocity in increased.


\begin{figure}
        \centering
        \includegraphics[width=1\linewidth]{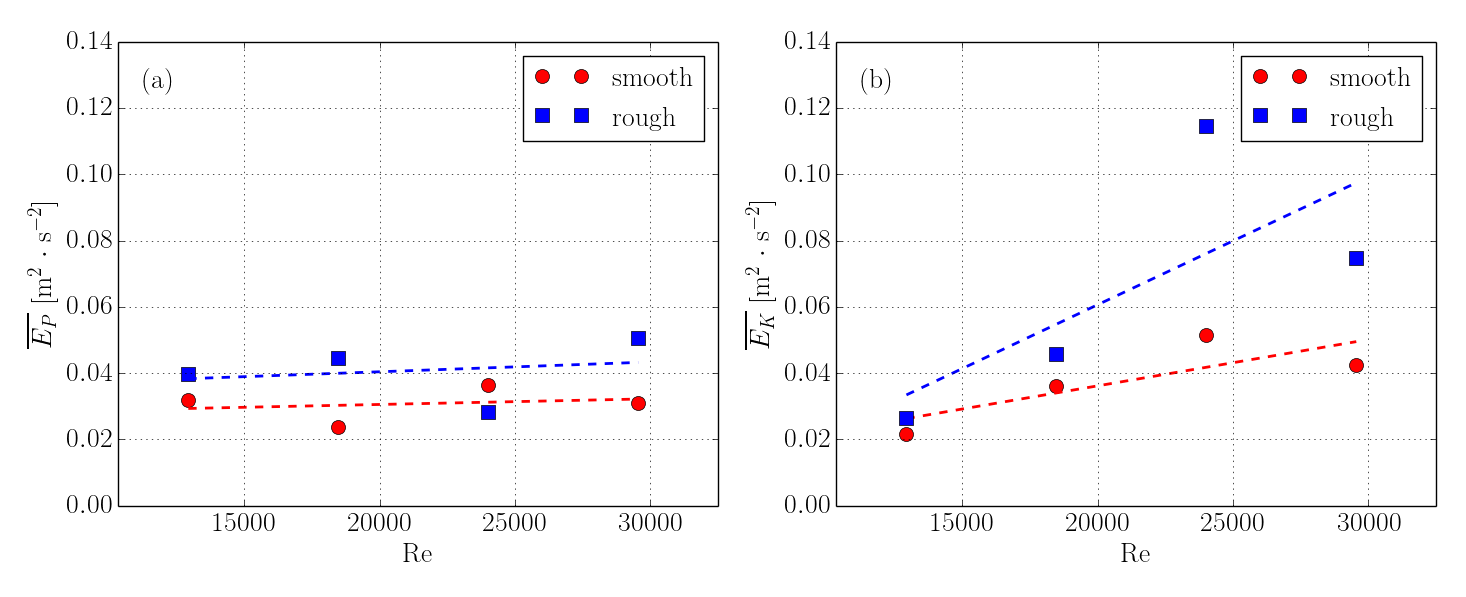}
        \caption{\label{fig:kinetic_potential_energy} Mean values of (a) potential and (b) kinetic energies for the different runs of the experiment. Symbols represent the measured values and dashed lines are trend lines added for clarity. \label{fig:energy_vs_Re}}
\end{figure}

With regard to the particle total energy, it is instructive to visualize the distributions of the components, namely the kinetic and potential energy of the particles. The mean energy components are presented in figure~\ref{fig:kinetic_potential_energy} with respect to the previously defined Reynolds number. From these graphs, we understand that the change in total energy with the motor velocity is solely due to changes in particle kinetic energy, 
while the average particle potential energy remains independent of the angular velocity of the vortex, yet higher for the rough surfaces (suggesting higher resuspension efficiency). Moreover, the values of particle energy are higher above the rough surface for all cases. In addition, as the Reynolds number grows, the kinetic energy level rises with higher rate for the rough surfaces, meaning that more energy is transferred to the resuspended particles. The decline of growth for the highest tested Reynolds number is also seen in this result. The effect is again due to higher centrifugal forces as the motor speed has increased.

A more detailed view is shown in the following figure~\ref{fig:horizontal_vertical_energy}, in which we demonstrate the horizontal and vertical (against gravity) particle kinetic energy components. Figure~\ref{fig:horizontal_vertical_energy}(a) shows the distribution of velocities magnitude in the vertical and the horizontal directions for the 100 rpm motor velocity run over the smooth and the rough surfaces, while figure \ref{fig:horizontal_vertical_energy}(b) shows the values of the mean particle kinetic energy in the vertical and the horizontal directions for increasing Reynolds numbers. Naturally, due to the type of the vortex flow, the particle kinetic energy is an order of magnitude larger in the horizontal relative to the vertical directions. It is also evident that over the rough surface the particles exhibit higher energy levels, with the trend similar for both vertical and horizontal components of energy. Moreover, the mean values of the velocity rise with the angular velocity of the vortex, again only up to 130 rpm. Although this behavior exists in both directions it is much more prominent in the horizontal direction.


\begin{figure}[ht]
\centering
    \includegraphics[width=1\linewidth]{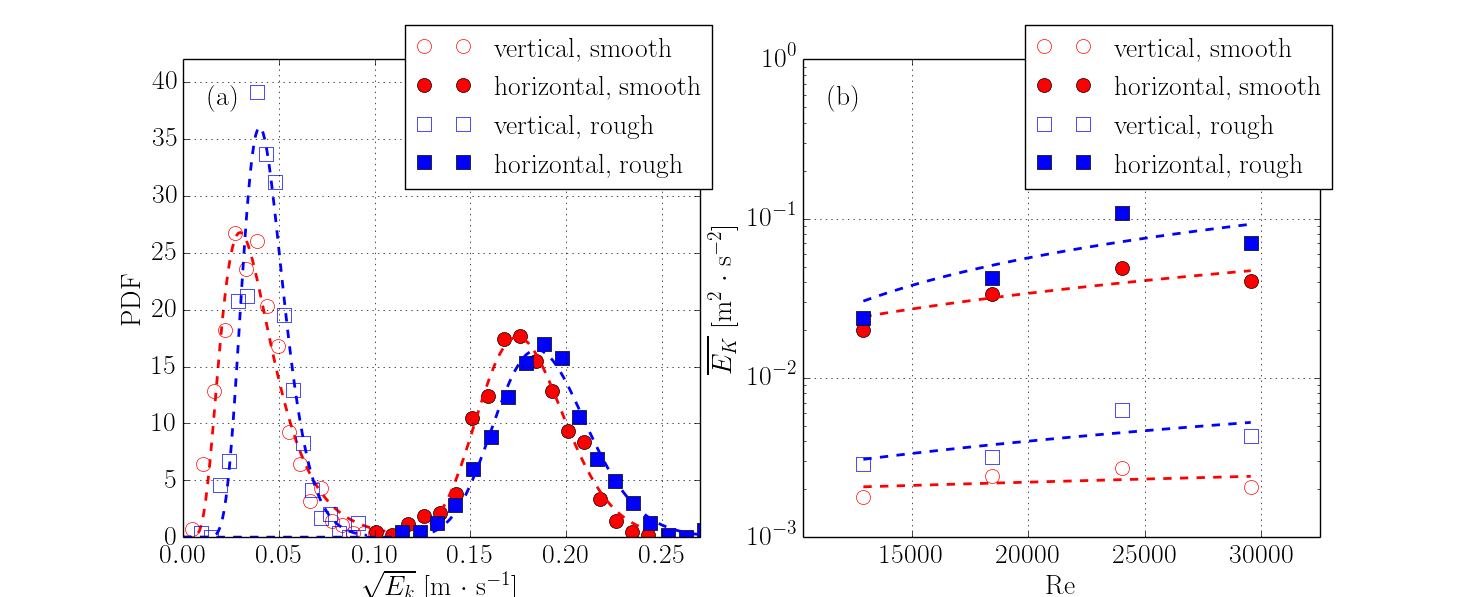}
\caption{(a) kinetic energy distributions in the horizontal and the vertical directions for the 100 RPM motor velocity run over the smooth and the rough surfaces (b) mean kinetic energy in the vertical and horizontal directions over the smooth and the rough surfaces for the different values of motor velocity \label{fig:horizontal_vertical_energy}}
\end{figure}

\subsection{Particle accelerations}

\begin{figure}[ht]
\centering
  \centering
  \includegraphics[width=1\linewidth]{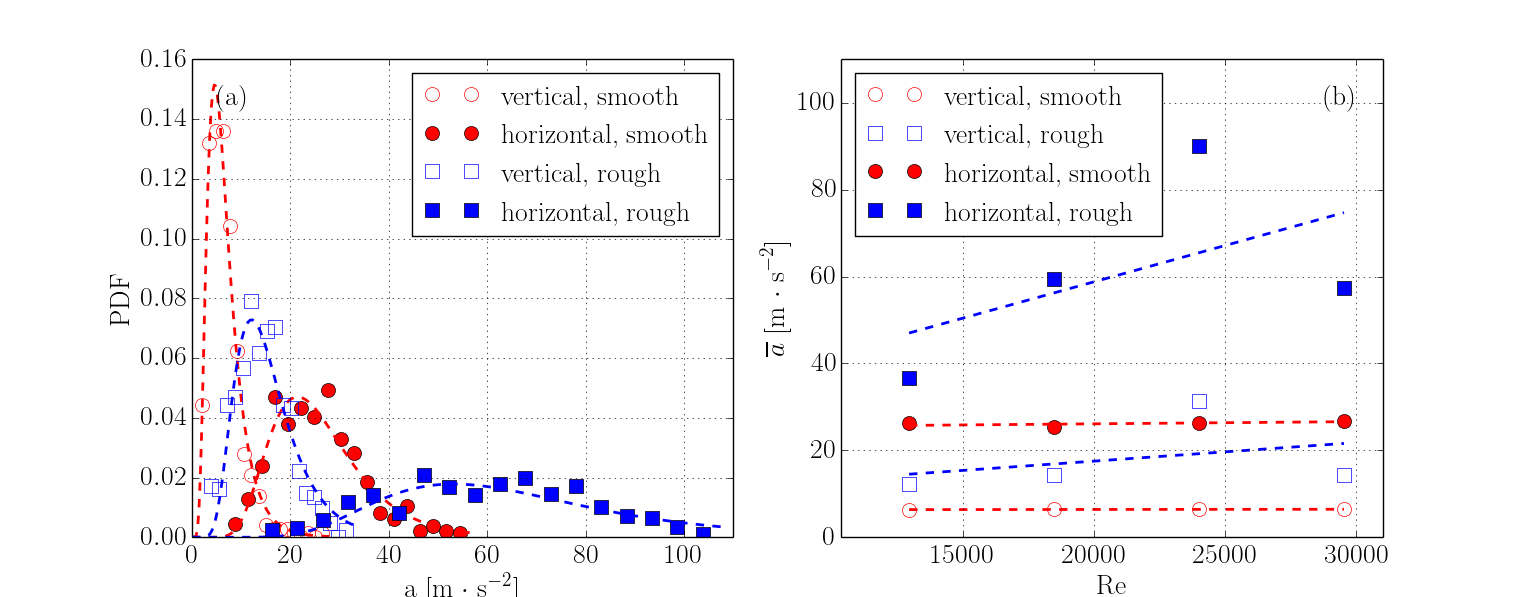}
\caption{Acceleration PDF in the horizontal and vertical directions for the smooth and the rough surfaces for the 100 RPM experiment; (b) mean particle accelerations in the horizontal and the vertical directions over the two different surfaces for the changing values of motor velocity \label{fig:accelerations}}
\end{figure}

3D-PTV allows us to analyze the particle material acceleration along trajectories in space and time, providing insight regarding the forces acting on the particles. Since the fluid velocity around the particles is unknown, it is impossible to decompose our results to the hydrodynamic forces of lift and drag. Nevertheless, it is useful to examine the particles accelerations and decompose it to the horizontal and the vertical components, since in the case of resuspension, the vertical component directly relates to the lift force, as long as the particle moves on the surface. Figure~\ref{fig:accelerations}(a) shows the acceleration probability distribution for the 100 RPM run over the smooth and the rough surfaces in the vertical and the horizontal components. Figure~\ref{fig:accelerations}(b) presents the mean acceleration values for the increasing Reynolds number over the two surfaces in the horizontal and the vertical directions. From these two graphs, it is obvious that in both directions and for all Reynolds numbers tested, the particles over the rough surface experience significantly higher accelerations, and thus much stronger forces. This is true for both vertical and horizontal components, but with values much higher in the horizontal 
(parallel to the wall) direction. The peculiar result is that there is almost no change in the accelerations for the particles over the smooth surface. It is also seen that although the horizontal accelerations (and centrifugal forces) increase with the increasing vortex speed, the vertical (up-lifting) forces are almost constant. This is consistent with the previous results of the vertical kinetic energy, and supports the assertion that the lift forces in this flow configuration are supported mostly by the pressure differences which are insensitive to the vortex angular velocity. Nevertheless, the important result for the purpose of this work is that both components of the acceleration are higher for the rough surface as compared to the smooth surface. This means that larger drag and lift forces are acting on the particles above the rough surface.

\subsection{Entrainment rate}

\begin{figure}[ht]
	\centering
	\includegraphics[width=.8\linewidth]{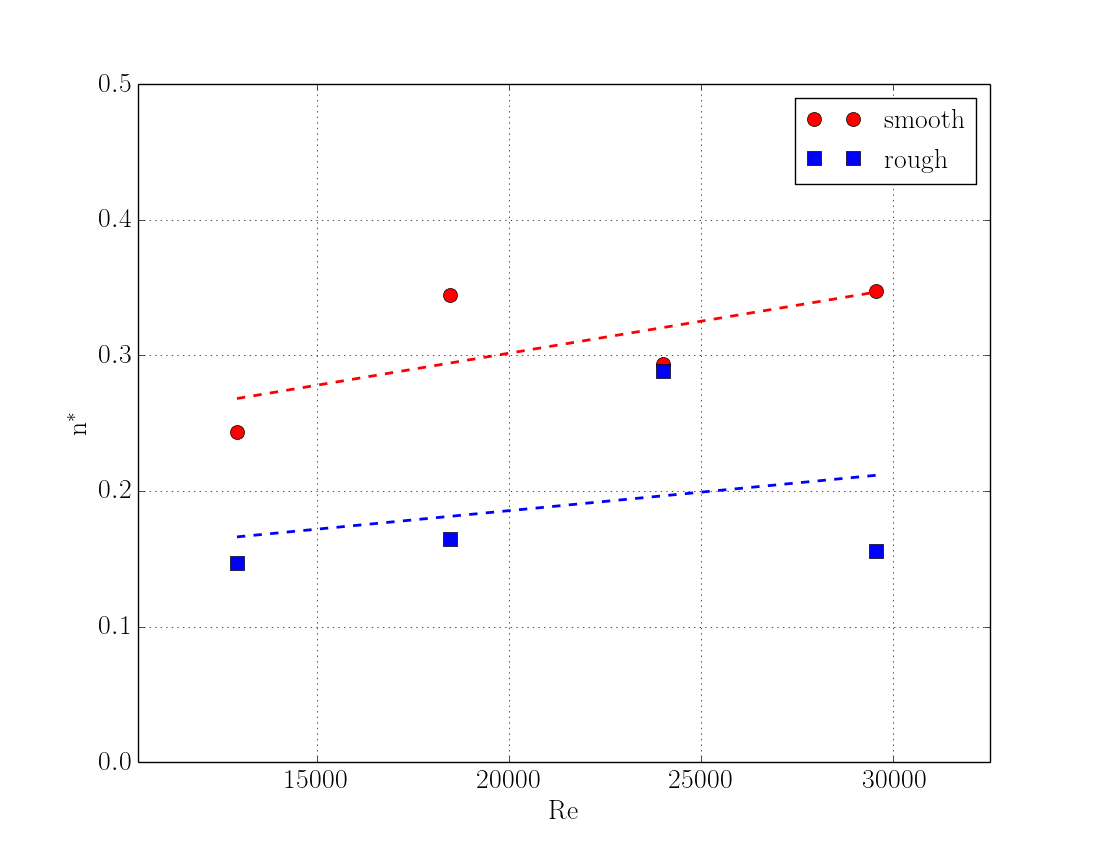}
	\caption{Fraction of particles remaining on the bottom wall for the different values of motor velocity over the smooth and the rough surfaces.\label{fig:fraction_remaining}}
\end{figure}

The most typical measure of a system resuspension efficiency is the fraction of particles that are entrained, or its complementary part, the fraction of particles that remain on the floor.
 This integral measure allows to quantify relative tendency for a given type of particles to become resuspended under different flow conditions. For each run of the experiment, the number of resuspended particles ($n^*$), and the total number of particles ($n_T$), were counted, and using the Eq.~\ref{eq:3} the fraction of the particles that remain on the smooth or rough surface was obtained. Figure \ref{fig:fraction_remaining} shows the fraction remaining for each experimental run as a function of the increasing Reynolds number of the flow. It is seen from the graph that the particles above the smooth surface are more likely to remain on the tank bottom wall for all the vortex velocities tested. It is also evident that with the increasing vortex angular speed, a slightly larger portion remains on the bed. This is explained by much higher centrifugal forces that remove particles from the vortex core, and thus decrease their chances to experience vertical lift forces which are strongest at the vortex core.
The trends also show some resemblance to the mean potential energy curves plotted in figure \ref{fig:energy_vs_Re}a. In examining the differences between figures \ref{fig:fraction_remaining} and \ref{fig:energy_vs_Re}a, we infer that while the potential energy levels remain roughly the same above the two surfaces, it is clear that the number of resuspended particles is higher above the rough surface. This means that above the rough surface more particles are resuspended as compared to the smooth surface case, but to slightly lower heights. It is again linked to the higher horizontal accelerations that the particles experience above the rough surfaces. The following discussion section is devoted to a simplified mechanistic model that explains the differences between the two types of surfaces.

\section{Summary and discussion}

In this work, the 3D-PTV was used to study the effect of surface roughness  on the different modes of resuspension. The study focused on the detailed motion analysis of single solid particles from smooth and rough surfaces. In a quasi static vortex flow, the energy levels, the velocities, and the amount of resuspended particles were measured. The results were obtained through a 3D-PTV system in a simple vortex flow that provided us with the detailed information
regarding the positions, velocities, and accelerations of particles to be collected for a long period of time. The results presented in this work show unequivocally that in the case of spherical particles with a diameter larger than the
viscous sublayer, roughness improves the resuspension rate. It is important to note that this result is new both at the level of the detailed information about trajectories of the freely moving particles as they are suspended or deposited, and at the direct comparative study of the smooth versus rough surfaces under the same flow conditions. This experiment allows us to single out the mechanism by which the resuspension rate increases for the mobile beds over the rough surfaces.

\begin{figure}[!ht]
\centering
\begin{tikzpicture}
\node[inner sep=0pt] (image) at (0,0)
    {\includegraphics[width=.8\textwidth]{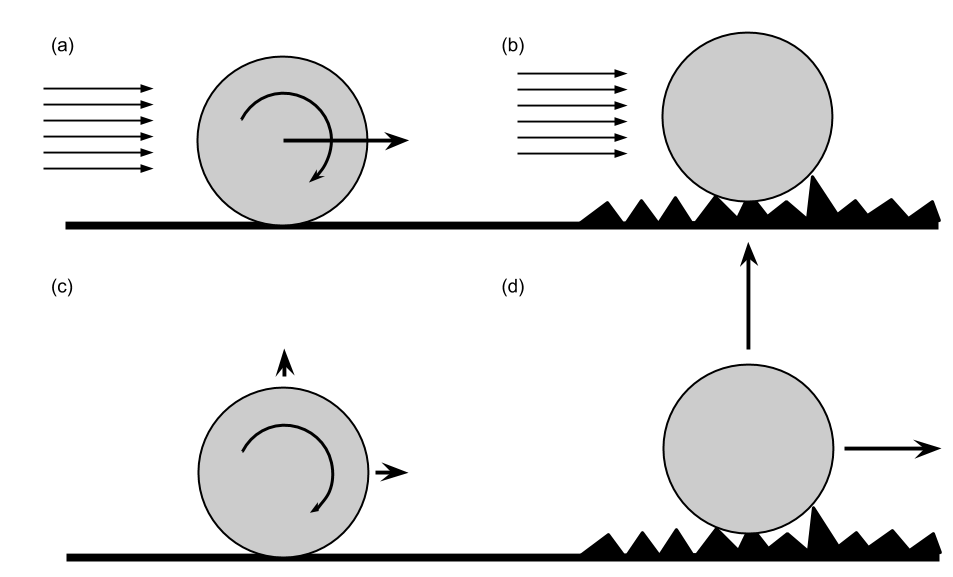}};
\node at (-.5,2) {$V_p$};
\node at (-5,3) {$U$};
\node at (1.2,3.2) {$U$};
\node at (-2.5,-.5) {$L$};
\node at (3.25,.5) {$L$};
\node at (-.75,-2.5) {$D$};
\node at (6.5,-2) {$D$};
 \end{tikzpicture}
\caption{Cartoon of the large spherical particle resuspension from smooth (left panel) versus rough (right panel) surfaces: (a) spherical particle under the flow from the left can roll or slide at the velocity comparable to the flow velocity, (b) same particle on the rough surface cannot roll or slide due to the roughness, (c) the very small relative velocity of the particle decreases significantly the drag and lift forces responsible for the resuspension; (d) lift and drag forces are enhanced due to a very high relative velocity which in this model is equal to the flow velocity.\label{fig:cartoon}}
\end{figure}
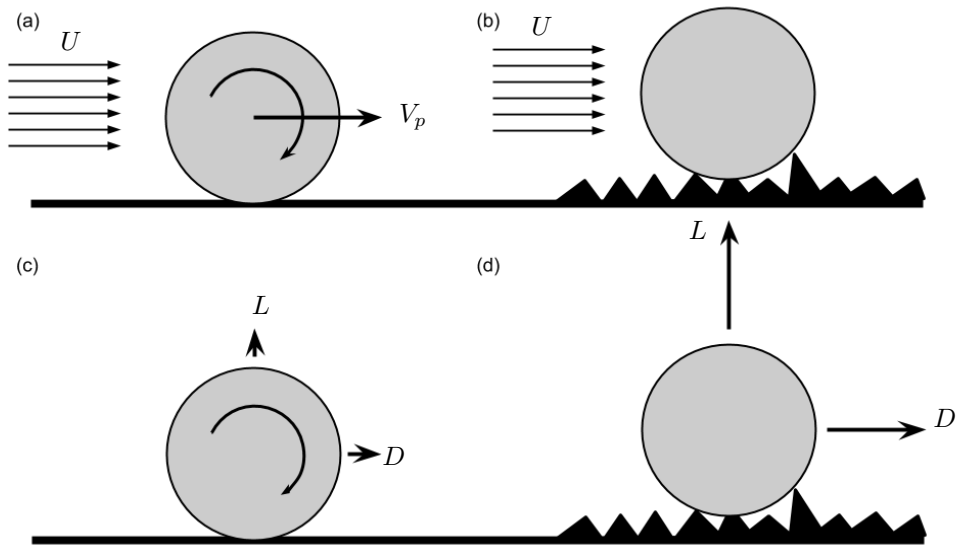

We would like to propose a simplified conceptual mechanistic model that explains the differences of the spherical particles resuspension from smooth and rough surfaces. The model is relevant only for the particles large enough, such that the adhesion to the smooth surface is negligible, as compared to hydrodynamic forces. In this case, the spherical particle on a smooth surface can roll or slide, whereas on a rough surface this ability is impaired. A conceptual cartoon is shown in figure~\ref{fig:cartoon}. The central concept is the ability of the particles to roll/slide on the surface. The major difference and the increased resuspension rate over the rough surface is therefore due to the strongly increased relative velocity between the moving particle and the flow, i.e. $W = U - V_p$ ( figure~\ref{fig:cartoon} top panel). 
Since the major hydrodynamic forces of lift and drag are both proportional to some power of the relative velocity $L,D \propto W^\alpha$, 
which, depending on the particle Reynolds number, 
can be in the range of $\alpha = 1 \div 2$, the increase of the forces can be up to the $W^2$.

As one can see from our results of resuspension efficiency (figure~\ref{fig:fraction_remaining}), velocity and acceleration of the particles that are lifted up to the flow (figure~\ref{fig:kinetic_potential_energy}),  these resemble the effects predicted by this mechanistic model. Moreover, the model also predicts that at some energy level, the roughness cannot sustain the energy transferred to the particle by the flow, and thus the effect of increasing suspension rate is less prominent. In our case, this happened at the highest rotation rate shown here of 160 rpm, or $Re \sim 30000$. The higher rotation rates have shown the same trend - the velocities were too high and as the particle mobility along the wall was recovered,
the resuspension rate decreased towards the smooth wall case. 
 Also similar results concerning particle mobility over the surface were observed by \cite{Ibrahim:2003} where it was
 shown that particles with a less spherical shape were more easily resuspended than their perfectly spherical counterparts (with higher degree of mobility).

In further support of our mechanistic model of strongly different forces acting on the particles, where mobility along the wall is hindered on the rough surfaces, we provide the direct measurements of the horizontal accelerations of the conditionally sampled subset of particles where their centers are within one radius from the wall. The results of the horizontal accelerations, i.e. forces, are shown in the following figure~\ref{fig:horizontal_acceleration}. Obviously, the distributions of the forces experienced by the particles on the rough surfaces are several times higher than those on the smooth surface, 
despite the exact same flow and other conditions of the experiments.  
 We see this result as a direct support of the proposed model. It is also clear that the effect saturates at high velocities, as predicted by the model.

\begin{figure}[ht]
	\centering
	\includegraphics[width=.8\linewidth]{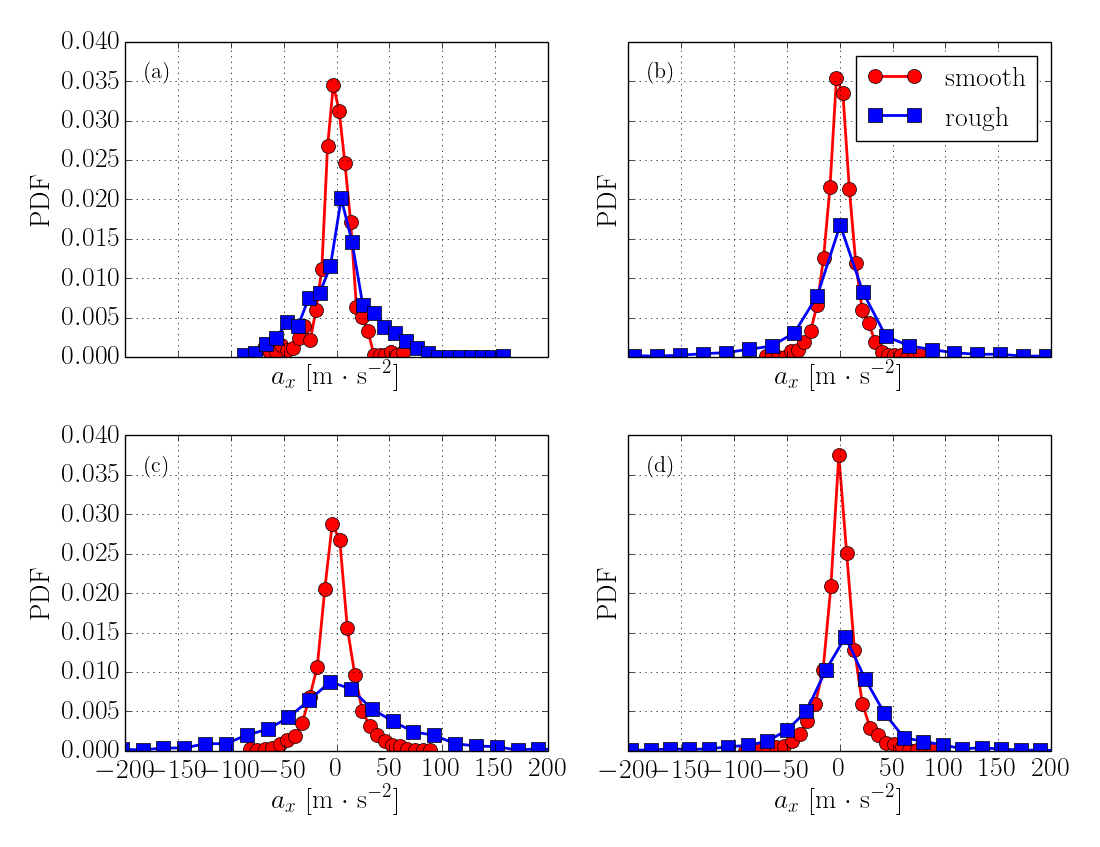}
	\caption{Horizontal acceleration component of the particles on smooth and rough surfaces.}
	\label{fig:horizontal_acceleration}
\end{figure}

We also would like to discuss the reasons for differences found in the results of several studies concerning the effect of surface roughness.
\cite{nino:2003} and \cite{lee:2012} have concluded that surface roughness increases the threshold shear velocity for particle resuspension (and reducing resuspension efficiency). It is important to note that in those studies the surface roughness size was taken to be at the order of the size of the particle diameter, such that the surface elements  are hiding the particles from the flow.
 Another study by \cite{Soltani:1995} showed lower resuspension threshold for particles over a surface with roughness much smaller than the particle diameter.
These different effects can be seen in the results of \cite{jiang:2008}, concluding that surface roughness affects particles of different relative (to the surface roughness and to the viscous sub-layer thickness) diameters in different ways.

The results of this study add to the existing knowledge regarding the resuspension mechanisms of freely moving particles over smooth and rough surfaces. The new results, based on the detailed measurements of the resuspended particles, allow for a better modeling and analysis of suspensions in industrial applications of fluidized beds, bio-reactors, mixers of solids, and environmental problems of dust devils or tornadoes. Beyond the flow types that resemble the vortex model used in this study, we can understand better the observations of strong, stochastic suspension events in the turbulent boundary layers. In addition to the commonly accepted role of sweeps and ejections, we present a simple mechanistic model based on the inhomogeneous roughness of the surface that can provide an alternative and simpler explanation of the irregular and abrupt resuspension events.


\bibliographystyle{apalike2}

\begin{thebibliography}{}

\bibitem[Barth et~al., 2014]{Barth:2014}
Barth, T., Preuss, J., Muller, X., \& Hampel, X. (2014).
\newblock Single particle resuspension experiments in turbulent channel flows.
\newblock {\em Journal of Aerosol Science}, 71, 4--51.

\bibitem[Dracos, 1996]{Dracos:1996}
Dracos, T.~H. (1996).
\newblock {\em {Three-Dimensional Velocity and Vorticity Measuring and Image
  Analysis Technique: Lecture Notes from the short course held in Zurich,
  Switzerland}}.
\newblock Kluwer Academic Publisher.

\bibitem[Greeley et~al., 2004]{Greeley:2004}
Greeley, R., Whelley, P.~L., \& Neakrase, L. D.~V. (2004).
\newblock Martian dust devils: Directions of movement inferred from their
  tracks.
\newblock {\em Geophysical Research Letters}, 31.

\bibitem[Grzybowski \& Gradon, 2007]{Grzybowski:2007}
Grzybowski, K. \& Gradon, L. (2007).
\newblock Re-entrainment of particles from powder structure: experimental
  investigations.
\newblock {\em Advanced Powder Technology}, 18, 427--439.

\bibitem[Hall, 1988]{Hall:1988}
Hall, D. (1988).
\newblock Measurements of the mean force on a particle near a boundary in
  turbulent flow.
\newblock {\em Journal of Fluid Mechanics}, 187, 451--466.

\bibitem[Henry \& Minier, 2014]{henry_minier:2014}
Henry, C. \& Minier, J.-P. (2014).
\newblock Progress in particle resuspension from rough surfaces by turbulent
  flows.
\newblock {\em Progress in Energy and Combustion Science}, 45, 1--53.

\bibitem[Henry et~al., 2012]{Henry:2012}
Henry, C., Minier, J.-P., \& Lefevre, G. (2012).
\newblock Numerical study on the adhesion and reentrainment of nondeformable
  particles on surfaces: The role of surface roughness and electrostatic
  forces.
\newblock {\em Langmuir}, 28, 438--452.

\bibitem[Ibrahim \& Dunn, 2003]{Ibrahim:2003}
Ibrahim, A. \& Dunn, P. (2003).
\newblock Microparticle detachment from surfaces exposed to turbulent air flow:
  controlled experiments and modeling.
\newblock {\em Aerosol Science}, 34, 765--782.

\bibitem[Jiang et~al., 2008]{jiang:2008}
Jiang, Y., Matsusaka, S., Masuda, H., \& Qian, Y. (2008).
\newblock Characterizing the effect of substrate surface roughness on
  particle‒wall interaction with the airflow method.
\newblock {\em Powder Technology}, 186(3), 199--205.

\bibitem[Lee \& Balachandar, 2012]{lee:2012}
Lee, H. \& Balachandar, S. ({2012}).
\newblock {Critical shear stress for incipient motion of a particle on a rough
  bed}.
\newblock {\em Journal of Geophysical Research - Earth Surface}, {117}.

\bibitem[Nino et~al., 2003]{nino:2003}
Nino, Y., Lopez, F., \& Garcia, M. ({2003}).
\newblock {Threshold for particle entrainment into suspension}.
\newblock {\em Sedimentology}, {50}({2}), {247--263}.

\bibitem[Soltani \& Ahmadi, 1995]{Soltani:1995}
Soltani, M. \& Ahmadi, G. (1995).
\newblock Particle detachment from rough surfaces in turbulent flows.
\newblock {\em Adhesion}, 51, 105--123.

\bibitem[Traugott et~al., 2011]{Traugott:2011}
Traugott, H., Hayse, T., \& Liberzon, A. (2011).
\newblock Resuspension of particles in an oscillating grid turbulent flow using
  piv and 3d-ptv.
\newblock {\em Journal of Physics: Conference Series}, 318.

\bibitem[Wu \& Chou, 2013]{Wu:2013}
Wu, F. \& Chou, Y. (2013).
\newblock Rolling and lifting probabilities for sediment entrainment.
\newblock {\em Journal of Hydraulic Engineering}, 129, 110--119.

\bibitem[Yanbin et~al., 2008]{YanbinJiang:2008}
Yanbin, J., Matsusaka, S., Masuda, H., \& Qian, Y. (2008).
\newblock Characterizing the effect of substrate surface roughness on
  particle--wall interaction with the airflow method.
\newblock {\em Powder Technology}, 186, 199--205.

\bibitem[Ziskind, 2006]{Ziskind:2006}
Ziskind, G. ({2006}).
\newblock {Particle resuspension from surfaces: Revisited and re-evaluated}.
\newblock {\em Reviews in Chemical Engineering}, {22}({1-2}), {1--123}.

\end{thebibliography}

\end{document}